\documentclass[aps]{revtex4}
\usepackage{epsf}
\usepackage{bm}
\newcommand{\tjs}[6]{\left(
\begin{array}{ccc}
#1& #2& #3\\ #4& #5& #6
\end{array}\right)}
\newcommand{\sjs}[6]{\left\{
\begin{array}{ccc}
#1& #2& #3\\ #4& #5& #6
\end{array}\right\}}

\begin{document}

\title{QED Calculation of E1M1 and E1E2 Transition
Probabilities in One-Electron Ions with Arbitrary Nuclear Charge.}

\author{L.N.Labzowsky${}^{1,2}$, A.V.Shonin${}^{1}$ and D.A.Solovyev${}^{1}$\\
1) Institute of Physics, St Petersburg State University, 198904,
Uljanovskaya 1, Petrodvorets, St Petersburg, Russia\\ 2)
Petersburg Nuclear Physics Institute, 188350, Gatchina, St
Petersburg, Russia\\}

\begin{abstract}
The quantum electrodynamical (QED) theory of the two-photon
transitions in hydrogenlike ions is presented. The emission
probability for $2s_{1/2}\rightarrow 2\gamma(E1)+1s_{1/2}$
transitions is calculated and compared to the results of the
previous calculations. The emission probabilities
$2p_{1/2}\rightarrow\gamma(E1)+\gamma(E2)+1s_{1/2}$ and
$2p_{1/2}\rightarrow\gamma(E1)+\gamma(M1)+1s_{1/2}$ are also
calculated for the nuclear charge $Z$ values $1\leq Z\leq 100$.
This is the first calculation of the two latter probabilities. The
results are given in two different gauges.
\end{abstract}

\maketitle

\begin{center}
\textbf{1.Introduction.}
\end{center}
The probabilities of the two-photon spontaneous decay in the
hydrogen atoms and hydrogenlike ions were studied since the
theoretical formalism was worked out by Goeppert-Mayer~\cite{q1}
and the first estimate for the two-photon transitions
$2s_{1/2}\rightarrow 2\gamma(E1)+1s_{1/2}$ was obtained by Breit
and Teller~\cite{q2}. This transition in H atom and low-Z H-like
ions is of special importance since it defines the lifetime of
$2s$ level and exceeds by many orders of magnitude, approximately
by $(1/{\alpha Z})^4$, the probability of the one-photon decay
$2s\rightarrow \gamma(M1)+1s$. Here $\alpha$ is the fine structure
constant, $Z$ is the charge of the nucleus. A highly accurate
nonrelativistic calculations of the transition probability for
$2s\rightarrow 2\gamma(E1)+1s$ process for the hydrogen atom was
performed by Klarsfeld~\cite{q3}. The first fully relativistic
calculation of this transition probability was made by
Johnson~\cite{q4} and later by Goldman and
Drake~\cite{q5},~\cite{q6}, and by Parpia and Johnson~\cite{q7}.
The recoil corrections were given in the papers by Fried and
Martin~\cite{q8} and by Bacher~\cite{q9}. More recently,
Karshenboim and Ivanov~\cite{q9a} evaluated the radiative
corrections to this decay. The decay probabilities $ns\rightarrow
2\gamma(E1)+1s$ with $n=3-6$ in H atom were evaluated
in~\cite{q9b}.

The E1M1 two-photon decay rate was so far evaluated only for the
transition $2^3P_0\rightarrow\gamma(E1)+\gamma(M1)+1^1S_0$ in
two-electron ions. The reason is that for the level $2^3P_0$ this
decay channel is dominant in the absence of the hyperfine
quenching as was first stated in~\cite{q10}. According
to~\cite{q10}, the angular momentum coupling rules for
$0\rightarrow 0$ transitions allow only the emission of two
photons with equal values of the angular momentum. The probability
of two-photon decay
$2^3P_0\rightarrow\gamma(E1)+\gamma(M1)+1^1S_0$ was evaluated
within fully relativistic approach for $Z=92$ in~\cite{q11} and
for $50\leq Z \leq 94$ in~\cite{q12}. Recently a rigorous QED
approach~\cite{q12a} was applied to the evaluation of E1M1
transition in He-like ions with $30\leq Z\leq 92$~\cite{q12b}.

In this paper we describe the QED theory of two-photon decay
process and calculate the $E1M1$ and $E1E2$ transition
probabilities for H-like ions for nuclear charge $Z$ values within
the region $1\leq Z\leq 100$.

Unlike the case of two-electron ions, both E1M1 and E1E2
transitions are allowed in the one-electron systems. As far as we
know the transitions $2p\rightarrow\gamma(E1)+\gamma(M1)+1s$ and
$2p\rightarrow\gamma(E1)+\gamma(E2)+1s$ were never calculated. The
reason is, of course, that they are very small (about $10^{-13}$
for $Z=1$) compared to the leading transition
$2p\rightarrow\gamma(E1)+1s$. Still both two-photon transitions
behave like $Z^8$ for larger $Z$ values. The Z behavior of E1
transition is essentially weaker ($Z^4$) and for $Z=92$ the E1
transition prevails only by two orders of magnitude.

We performed the calculation in two different gauges, thus
receiving an accurate check of the gauge invariance of the
results. The two gauges employed were relativistic counterparts of
the "velocity" and "length" forms of the transition amplitudes in
the nonrelativistic case. In the low-Z region in the "velocity"
gauge the intermediate negative-energy states give nearly dominant
contributions both to E1M1 and E1E2 transition probabilities.
Contrary to this, the negative-energy states contribution is fully
negligible in the "length" gauge for the low-Z values. Similar
conclusions on the importance of the negative-energy states in
"velocity" gauge for the calculations of the one-photon transition
probabilities in neutral atoms with one valence electron within
the Relativistic Many Body Perturbation Theory approach were made
earlier by Savukov and Johnson~\cite{q12c}.

In this paper we derive also an explicit expression for the
negative-energy contribution to the decay probabilities E1M1 and
E1E2 for the low-Z H-like ions in "velocity" gauge.

In the QED calculation of the two-photon transition probabilities
one needs to generate the complete Dirac spectrum. In this paper
we used the Dirac-Coulomb wave functions. For the summation over
complete Dirac spectrum the different powerful numerical methods
were developed in the last decades: the finite basis set
method~\cite{q13}, the $B$-spline method~\cite{q14} and the space
discretization method~\cite{q15}. In this work we used the version
of the B-spline method presented in~\cite{q16}. The relativistic
units are used throughout this paper.

\begin{center}
\textbf{2. QED theory of two-photon transitions.}
\end{center}

The two-photon decay process $A\to A'+2\gamma$ for the
noninteracting electrons is represented by the Feynman graphs
Fig.1. In this section we characterize the photons by the momentum
$\textbf{k}$ and the polarization $\textbf{e}$. According to the
Feynman correspondence rules the S-matrix element $S^{(1a)}_{A'A}$
is equal to~\cite{q17},~\cite{q18}.
\begin{equation}
\label{u7}
\begin{array}{l}
\displaystyle S^{(1a)}_{A'A}=(-i)^2e^2\int
d^4x_1d^4x_2\left(\bar\Psi_{A'}(x_1)\gamma_{\mu1}
A^*_{\mu1}(x_1)S(x_1x_2)\gamma_{\mu2}A^*_{\mu2}(x_2)\Psi_A(x_2)\right)
\end{array}
\end{equation}
\noindent Here

\begin{equation}
\label{u8} \displaystyle S(x_1x_2)=\frac{1}{2\pi
i}\int\limits_{-\infty}^{\infty} d\omega_1
e^{i\omega_1(t_1-t_2)}\sum_n \frac{\Psi_n({\textbf{x}_1})
\bar\Psi_n(\textbf{x}_2)}{E_n(1-i0)+\omega_1}
\end{equation}
\noindent is the electron propagator where the sum runs over the
Dirac spectrum for the electron in the field of the nucleus,

\begin{equation}
\label{u9}
\begin{array}{l}
\displaystyle\Psi_A(x)=\Psi_A(\textbf{r})e^{-iE_At}
\end{array}
\end{equation}
\noindent is the electron wave function, $E_A$ is the electron
energy,

\begin{equation}
\label{u10}
\begin{array}{l}
\displaystyle
A^{\textbf{k},\lambda}_{\mu}(x)=\sqrt{\frac{2\pi}{\omega}}
e^{(\lambda)}_{\mu}e^{i(\textbf{k}\textbf{r}-\omega t)}
\end{array}
\end{equation}
\noindent is the wave function of the photon characterized by the
momentum $\textbf{k}$ and polarization vector $e_{\mu}^{\lambda}$
($\mu,\lambda=1,2,3,4$), $x\equiv(\textbf{r},t)$. For the real
transverse photons

\begin{equation}
\label{u28}
\begin{array}{l}
\displaystyle\textbf{A}(x)=\sqrt{\frac{2\pi}{\omega}}\textbf{e}
e^{i(\textbf{k}\textbf{r}-\omega t)}\equiv
\sqrt{\frac{2\pi}{\omega}}\textbf{A}_{\textbf{e},\textbf{k}}
(\textbf{r})e^{-i\omega t}
\end{array}
\end{equation}
Inserting Eqs.~(\ref{u8})-(\ref{u28}) in
Eq.~(\ref{u7}),integrating over time and frequency variables and
introducing the amplitude $U_{A'A}$ as

\begin{equation}
\begin{array}{l}
\displaystyle S_{AA'}=-2\pi
i\delta(E_{A'}+\omega+\omega'-E_A)U_{A'A}
\end{array}
\end{equation}
\noindent we obtain

\begin{equation}
\label{u11} \displaystyle U^{(1a)}_{A'A}=\frac{2\pi
e^2}{\sqrt{\omega\omega'}}
\sum_n\frac{({\bm\alpha}\textbf{A}^*_{\textbf{e},
\textbf{k}})_{A'n}({\bm\alpha}\textbf{A}^*_{\textbf{e}',
\textbf{k}'})_{nA}}{E_n-E_A+\omega'}.
\end{equation}
where $e=\sqrt{\alpha}$ is the electron's charge.

Defining the transition probability as
\begin{equation}
\label{u12} \displaystyle
dW_{A'A}=2\pi\delta(E_A-E_{A'}-\omega-\omega')\left|U^{1a}_{A'A}+U^{1b}_{A'A}
\right|^2\frac{d\textbf{k}}{(2\pi)^3}\frac{d\textbf{k}'}{(2\pi)^3}
\end{equation}
\noindent and integrating over $\omega$ we obtain finally
\begin{equation}
\label{u13}
\begin{array}{l}
\displaystyle dW_{A'A}(\omega',{\bm{\nu}},{\bm{\nu}}',
\textbf{e},\textbf{e}')=e^4
\frac{\omega'(E_A-E_A'-\omega')}{(2\pi)^3}\nonumber\\
\displaystyle
\left|\sum_{n}\frac{({\bm{\alpha}}\textbf{A}^*_{\textbf{e},
\textbf{k}})_{A'n} ({\bm{\alpha}}\textbf{A}^*_{\textbf{e}',
\textbf{k}'})_{nA}}{E_n-E_A+\omega'}+\right.\nonumber\\
\displaystyle\left.\sum_{n}\frac{({\bm{\alpha}}\textbf{A}^*_{\textbf{e}',
\textbf{k}'})_{A'n} ({\bm{\alpha}}\textbf{A}^*_{\textbf{e},
\textbf{k}})_{nA}}{E_n-E_A+\omega}\right|^2d{\bm{\nu}}
d{\bm{\nu}}'d\omega'
\end{array}
\end{equation}
\noindent where ${\bm{\nu}}\equiv\textbf{k}/\omega$

It is more convenient to come over to the photon's wave functions
characterized by the angular momentum and parity. For this purpose
we will use the expansion of the linearly polarized wave in
spherical harmonics~\cite{q17},~\cite{q18}:
\begin{equation}
\label{u131} {\textbf{e}}
e^{i({\textbf{k}}{\textbf{r}})}=\sum_{J,M,\lambda=0,1}{\textbf{A}}^{(\lambda)}_{JM}({\textbf{r}})\left({\textbf{e}}{\textbf{Y}}^{(\lambda)}_{JM}({\textbf{k}})\right)
\end{equation}
where ${\textbf{A}}^{(\lambda)}_{JM}({\textbf{r}})$ is the
electromagnetic vector potential,
${\textbf{Y}}^{(\lambda)}_{JM}({\textbf{k}})$ is the vector
spherical function of the magnetic ($\lambda=0$) or electric
($\lambda=1$) type. The electric and magnetic vector potentials
are:
\begin{equation}
\label{u132}
\begin{array}{l}
\displaystyle
 \textbf{A}^{(1)}_{JM}(\textbf{r})=
i^{J+1}\left\{\sqrt{\frac{J}{2J+1}}g_{J+1}(kr)\textbf{e}{\textbf{Y}}^*_{JJ+1M}(\textbf{r})-\sqrt{\frac{J+1}{2J+1}}g_{J-1}(kr)\textbf{e}{\textbf{Y}}^*_{JJ-1M}(\textbf{r})+\right.\\
\displaystyle \left.G_J\left(
\sqrt{\frac{J+1}{2J+1}}g_{J+1}(kr)\textbf{e}{\textbf{Y}}^*_{JJ+1M}(\textbf{r})+\sqrt{\frac{J}{2J+1}}g_{J-1}(kr)\textbf{e}{\textbf{Y}}^*_{JJ-1M}(\textbf{r})+ig_{J-1}(kr)Y^*_{JM}\right)\right\}
\end{array}
\end{equation}

\begin{equation}
\label{u133} \textbf{A}^{(0)}_{JM}(\textbf{r})=
i^{J}g_{J}(kr)\textbf{e}{\textbf{Y}}^*_{JJM}(\textbf{r}).
\end{equation}

Here $\displaystyle g_{L}(\omega
r)=(2\pi)^{3/2}\frac{1}{\sqrt{\omega r}}J_{L+1/2}(\omega r)$,
$J_n$ is the Bessel function, ${\textbf{Y}}_{JLM}(\textbf{k})$ are
the vector spherical functions, $J, M$ are the photons angular
momentum and its projection, $G_J$ is gauge parameter  defining
gauge for the electromagnetic potentials. In our calculations we
employ the "velocity" gauge ($G_J=0$) and the "length" gauge
($\displaystyle G_J=\sqrt{\frac{J+1}{J}}$). Note, that
Eq.~(\ref{u28}) corresponds to $G_J=0$.

After substitution of Eqs.~(\ref{u131}-\ref{u133}) in
Eq.~(\ref{u13}) we can perform the summation over polarizations
and integration over the photon's angles using the formula
\begin{equation}
\label{u135} \sum_{e}\int d{\bm{\nu}}
({\textbf{e}}^*{\textbf{Y}}^{L}_{JM}({{\bm{\nu}}}))({\textbf{e}}{\textbf{Y}}^{L*}_{JM}({{\bm{\nu}}}))=\int
d{\bm\nu}({{\bm\nu}}\times{\textbf{Y}}^{L}_{JM}({{\bm\nu}}))
({{\bm{\nu}}}\times{\textbf{Y}}^{L'*}_{J'M'}({{\bm{\nu}}}))=\delta_{JJ'}\delta_{LL'}\delta_{MM'}
\end{equation}
Then the expression for the two-photon transition probability
looks like:
\begin{equation}
\label{u136}
\begin{array}{l}
\displaystyle dW_{A'A}(\omega')=e^4
\frac{\omega'(E_A-E_A'-\omega')}{(2\pi)^3}d\omega'\nonumber\times\\
\displaystyle \sum_{\lambda,M,J}\sum_{\lambda',M',J'}
 \left[\frac{({\bm\alpha}\textbf{A}^{(\lambda')*}_{J'M'\omega'})_
 {n_{A'}j_{A'}l_{A'}m_{A'},nj_{n}l_{n}m_{n}}
({\bm\alpha}\textbf{A}^{(\lambda)*}_
{JM\omega})_{nj_{n}l_{n}m_{n},n_{A}j_{A}l_{A}m_{A}}}
{E_{nj_n}-E_{n_Aj_A}+\omega}+\right.\nonumber\\
\displaystyle +\left.
 \frac{({\bm\alpha}\textbf{A}^{(\lambda)*}_{JM\omega})_
 {n_{A'}j_{A'}l_{A'}m_{A'},nj_{n}l_{n}m_{n}}
({\bm\alpha}\textbf{A}^{(\lambda')*}_
{J'M'\omega'})_{nj_{n}l_{n}m_{n},n_{A}j_{A}l_{A}m_{A}}}
{E_{nj_n}-E_{n_Aj_A}+\omega'}\right]\nonumber\\
\displaystyle
\end{array}
\end{equation}

Here we have replaced each electron subscript A by the standard
set of quantum numbers $n_A j_A l_A m_A$, where $n$ is the
principal quantum number, $j,m$ are the total electron angular
momentum and its projection and $l$ defines the parity of the
state.

In this work we calculate the total rate
\begin{equation}
\label{fw1} \displaystyle
W^{(2\gamma)}_{A'A}=\frac{1}{2}\int_0^{\omega_{max}}{\frac{dW_{A'A}}{d\omega'}}d\omega'
\end{equation}

where $\omega_{max}=E_A-E_{A'}$.

\begin{center}
{\textbf{ 3. The angular reduction}}
\end{center}

The angular integration in the matrix elements in Eq.~(\ref{u136})
can be performed in a standard way:
\begin{equation}
\label{f6}
\begin{array}{l}
\displaystyle({\bm\alpha}\textbf{A}^{(\lambda)*}_{
JM})_{n_{A'}j_{A"}l_{A'}m_{A'},n_{A''}l_{A''}j_{A''}m_{A''}}
=S_{j_{A'}m_{A'},j_{A''}m_{A''},J,M}
 C_{j_{A'}l_{A'},j_{A''}l_{A''},J}
 R^{(\lambda)}_{n_{A'}j_{A'},n_{A''}j_{A''},J}
\end{array}
\end{equation}

where

\begin{equation}
\begin{array}{l}
\displaystyle
S_{j_{A'}m_{A'},j_{A''}m_{A''},J,M}=(-1)^{-m_{A'}-M-1/2}
\tjs{j_{A'}}{j_{A''}}{J}{m_{A'}}{\bar m_{A''}}{M},
\end{array}
\end{equation}

the symbol $\bar m$  denotes $-m$, and

\begin{equation}
\begin{array}{l}
\displaystyle
C_{j_{A'}l_{A'},j_{A''}l_{A''},J,M}=\nonumber\\
{}\nonumber\\
\displaystyle =i\frac{1}{\sqrt{4\pi(2J+1)}}%
[j_{A'},j_{A''}]\tjs{j_{A'}}{J}{j_{A''}}{1/2}{0}{-1/2}\Pi^{(\lambda)}(l_{A'},l_{A''},J),
\end{array}
\end{equation}

\begin{equation}
\begin{array}{l}
\displaystyle \Pi^{(\lambda)}(l_{A'},l_{A''},J)=0 \mbox{ for odd }
(l_{A'}+l_{A''}+J+\lambda)\\
{}\\ \displaystyle \Pi^{(\lambda)}(l_{A'},l_{A''},J)=1 \mbox{ for
even } (l_{A'}+l_{A''}+J+\lambda),
\end{array}
\end{equation}

\begin{equation}
\begin{array}{l}
\displaystyle[j,j']=\sqrt{(2j+1)(2j'+1)}
\end{array}
\end{equation}
For the radial integrals we use the notations similar to ones
in~\cite{q11}:

\begin{equation}
\begin{array}{l}
\displaystyle R^{(0)}_{n_{A'}j_{A'},n_{A''}j_{A''},J}(\omega)=\sqrt{\frac{\omega}{2\pi}}\frac{2J+1}{\sqrt{J(J+1)}}%
(k_{A'}+k_{A''})I_J^+
\end{array}
\end{equation}

\begin{equation}
\label{f5}
\begin{array}{l}
\displaystyle R^{(1)}_{n_{A'}j_{A'},n_{A''}j_{A''},J}(\omega)=
\sqrt{\frac{\omega}{2\pi}}
\left[\sqrt{\frac{J}{J+1}}\left\{(k_{A'}-k_{A''})I_{J+1}^{+}+(J+1)I_{J+1}^{-}\right\}-\right.\nonumber\\
\displaystyle\left.\sqrt{\frac{J+1}{J}}\left\{(k_{A'}-k_{A''})I^{+}_{J-1}-JI^{-}_{J-1}\right\}+\right.\nonumber\\
\displaystyle\vphantom{\sqrt{\frac{J}{J+1}}}\left.G_J\left((2J+1)F_J+(k_{A'}-k_{A''})(I^{+}_{J+1}+I^{+}_{J-1})-JI^{-}_{J-1}+(J+1)I^{-}_{J+1}\right)\vphantom{\sqrt{\frac{J}{J+1}}}\right]
\end{array}
\end{equation}

\begin{equation}
\begin{array}{l}
\displaystyle I^{\pm}_{J}=\int r^2 dr g_J(\omega
r)(g_{A'}f_{A''}\pm f_{A'}g_{A''})
\end{array}
\end{equation}

\begin{equation}
\begin{array}{l}
\displaystyle F_{J}=\int r^2 dr g_J(\omega r)(g_{A'}g_{A''}+
f_{A'}f_{A''})
\end{array}
\end{equation}

Here $g_{nj}(r),f_{nj}(r)$ are the upper and lower radial
components of the Dirac wave function and

\begin{equation}
k=\left\{
\begin{array}{ccl}
\displaystyle l & \mbox{if} & j=l-\frac12\nonumber\\
\displaystyle -(l+1) & \mbox{if} & j=l+\frac12\\
\end{array}
\right.
\end{equation}

The total decay rate should be summed over the magnetic quantum
number $m_{A'}$ and averaged over the magnetic quantum number
$m_{A}$. Then

\begin{equation}
\label{f6a}
\begin{array}{l}
\displaystyle
dW_{A'A}(\omega')=\frac{e^4}{2\pi(2j_A+1)}d\omega'\times\\
\displaystyle\sum_{\lambda,J,\lambda',J'}\sum_{MM'}\sum_{m_{A'}m_{A}}\left[\sum_{nj_nl_n}
\frac{T^{(\lambda')}_{n_{A'}j_{A'}l_{A'},nj_nl_n,J'}(\omega')S_{j_{A'}m_{A'},j_nm_n,J'M'}
T^{(\lambda)}_{nj_nl_n,n_Aj_Al_A,J}(\omega)S_{j_nm_n,j_Am_A,JM}}{E_{nj_n}-E_{n_aj_a}+\omega}+\right.\\
\displaystyle\left.+(\omega\leftrightarrow\omega',
J\leftrightarrow J', \lambda\leftrightarrow\lambda',
M\leftrightarrow M')\vphantom{\frac{T^2}{T^2}}\right]^2
\end{array}
\end{equation}

where

\begin{equation}
\label{f7a} \displaystyle
T^{(\lambda)}_{n_Aj_Al_A,n_Bj_Bl_B,J}(\omega)\equiv
C^{(\lambda)}_{j_Al_A,j_Bl_B,J}R^{(\lambda)}_{n_Aj_{A'},n_Bj_B,J}(\omega)
\end{equation}

The summation over the $M, M', m_{A'}, m_A, m_n$ can be carried
 out using the sum rules for 3j-symbols:
\begin{equation}
\label{f8a} \displaystyle \sum_{\mbox{all
projections}}S_{j_{A'}m_{A'},j_nm_n, JM}S_{j_nm_n,j_Am_A, J'M'}
S_{j_{A'}m_{A'},j_{n'}m_{n'},JM}S_{j_{n'}m_{n'},j_Am_A,
J'M'}=\frac{\delta_{j_nJ_{n'}}}{2j_n+1}\, ,
\end{equation}

\begin{equation}
\label{f8b} \displaystyle \sum_{\mbox{all projections
}}S_{j_{A'}m_{A'},J_nm_n, JM}S_{j_nm_n,j_Am_A, J'M'}
S_{j_{A'}m_{A'},j_{n'}m_{n'},J'M'}S_{j_{n'}m_{n'},j_Am_A,
JM}=(-1)^{(J-J'+1)}\sjs{j_A}{J'}{j_n}{J_{A'}}{J}{j_{n'}}
\end{equation}

Inserting the expressions~(\ref{f6})-(\ref{f5}) in Eq.~(\ref{f6a})
and performing summations over all the projection indices we
obtain finally the formula for $dW_{A'A}$ expressed through the
various radial integrals:

\begin{equation}
\label{f9}
\begin{array}{l}
\displaystyle dW^{(2\gamma)}_{A'A}=e^4\frac{1}{2\pi(2j_A+1)}
\sum_{\lambda,\lambda',J,J'}(dW^{(1)}+dW^{(2)}+dW^{(3)}) d\omega'
\end{array}
\end{equation}

where

\begin{equation}
\label{f91} \displaystyle
dW^{(1)}=\sum_{j_n}\left[\frac{1}{4\pi(2j_n+1)}\sum_{nl_n}
\frac{T^{(\lambda)}_{n_{A'}j_{A'}l_{A'},nj_nl_n,J}(\omega)T^{(\lambda')}_{nj_nl_n,n_Aj_Al_A,J'}(\omega')}
{E_n-E_A+\omega'}\right]^2
\end{equation}

\begin{equation}
\label{f92} \displaystyle dW^{(2)}=dW^{(1)}
\,\,\,(\lambda\leftrightarrow\lambda', J\leftrightarrow J',
\omega\leftrightarrow\omega')
\end{equation}

\begin{equation}
\begin{array}{l}
\label{f93} \displaystyle
dW^{(3)}=\frac{1}{8\pi^2}\sum_{j_nj_{n'}}(-1)^{J-J'+1}\sjs{j_A}{J'}{j_n}{j_{A'}}{J}{j_{n'}}\\
\displaystyle
\sum_{nl_n}\frac{T^{(\lambda)}_{n_{A'}j_{A'}l_{A'},nj_nl_n,J}(\omega)T^{(\lambda')}_{nj_nl_n,n_Aj_Al_A,J'}(\omega')}
{E_n-E_A+\omega'}
\sum_{n'l_{n'}}\frac{T^{(\lambda')}_{n_{A'}j_{A'}l_{A'},nj_nl_n,J'}(\omega')T^{(\lambda)}_{nj_nl_n,n_Aj_Al_A,J}(\omega)}
{E_{n'}-E_A+\omega}
\end{array}
\end{equation}

\begin{center}
\textbf{4. $E1E1$ transition probabilities.}
\end{center}
The QED results of the calculations of the $E1E1$ two-photon
transition probabilities for H-like ions with nuclear charges
$Z=1\ldots 100$ in comparison with results from~\cite{q5} are
given in Table 1. For the summation over the entire Dirac spectrum
in Eqs.~(\ref{f91})-(\ref{f93}) the B-spline approach~\cite{q16}
was applied. The number of the grid points was $N=50$; the order
of splines $k=9$. The radial integration was performed by the
Gauss method with 10 integration points. Changing the number of
the grid points, the order of splines and the number of
integration points, we estimate our numerical inaccuracy as
$10^{-3}$. We solved the Dirac equations with the Fermi model for
the charge distribution $\rho(r)$ inside the nucleus:
\begin{equation}
\label{n33} \displaystyle \rho(r)=\frac{\rho_0}{exp[(r-c)/a]+1}
\end{equation}
with $a=0.5350$fm, $\rho_0$ is defined from the normalization
condition and $c$ deduced via the relation
$4\pi\int_0^{\infty}\rho(r)r^4dr=\langle r^2\rangle$. Here
$\langle r^2\rangle^{1/2}$ is the root-mean-square nuclear radius.

The results of our calculation agree well with the results of
calculation in~\cite{q5}. In the nonrelativistic limit the results
excibit the behavior
\begin{equation}
\label{n34} \displaystyle W^{E1E1}_{2s1s}=a^{E1E1}_{2s1s}(\alpha
Z)^6 a.u.
\end{equation}
with $a^{E1E1}_{2s1s}=0.001317$. This result coincides with the
accurate nonrelativistic value~\cite{q3}.

\begin{center}
\textbf{5. $E1M1$ transition probabilities.}
\end{center}
The numerical results for E1M1 $2p_{1/2}-1s_{1/2}$ transition
probabilities are given both in the "velocity" and "length" gauges
in Table 2. The details of the calculations are the same as for
the E1E1 transitions. The contributions of the positive-energy,
negative-energy and the "total" contributions are shown
separately. One has to remember, that the "total" contributions
includes also the interference contribution between the
positive-energy and negative-energy states.

According to the coupling rules for E1M1 $2p_{1/2}\rightarrow
1s_{1/2}$ decay the sets of the intermediate states in the sum in
Eq.~\ref{u136} are $ns_{1/2}, np_{1/2}, np_{3/2}$ and $nd_{3/2}$.
Of them, the states with $n=1,2$ give the dominant contribution
for small $Z$ values in the "length" gauge. This contribution
scales like
\begin{equation}
\label{n35} \displaystyle W^{E1M1 (+)}_{2p1s}(\rm{length})=W^{E1M1
(total)}_{2p1s}(\rm{length})=a^{E1M1}_{2p1s}(\alpha Z)^8 a.u.
\end{equation}
\begin{equation}
\label{n35a} \displaystyle \mbox{with} \phantom{qq}
a^{E1M1}_{2p1s}=2.907\cdot10^{-5}
\end{equation}

The negative-energy and the interference contributions are quite
negligible for small $Z$ values in the "length" gauge. The scaling
law Eq.~\ref{n35} can be understood from the estimate (in
relativistic units):

\begin{equation}
\label{n37} \displaystyle
W^{E1M1(+)}_{2p1s}(\rm{length})\sim\frac{\alpha^2}{\pi}\omega^7_{if}\left|\sum_{n(+)}\left\{\frac{\langle
i||d||n\rangle\langle n||{\mu}||f\rangle}{\Delta E_{ni}}+
\frac{\langle i||\mu||n\rangle\langle n||d||f\rangle}{\Delta
E_{ni}} \right\}\right|^2
\end{equation}

where $\omega_{if}$ is the transition frequency $2p-1s$, $\Delta
E_{ni}$ ar the effective energy denominators, $d$ and $\bm{\mu}$
are the electric and magnetic dipole transition operators,
$\langle a||\ldots||b\rangle$ are the reduced, i.e. integrated
over the angles matrix elements and the summation is extended over
the Schr\"{o}dinger equation solutions. The matrix elements for
the electric dipole operator $\textbf{d}=\sqrt{\alpha}\textbf{r}$
($\textbf{r}$ is the electron radius-vector in an atom) are of the
order $\langle i||d||n\rangle\sim\sqrt{\alpha}/{m\alpha Z}$ r.u.
The matrix elements of the magnetic dipole operator
${\bm{\mu}}=\sqrt{\alpha}{\mathbf{s}}/m$ ($\textbf{s}$ is the
electron spin) $\langle n||\mu||f\rangle$ are the order
$\sqrt{\alpha}/m$ if the principal quantum numbers for the
$\langle n|$ and $|f\rangle$ states coincide. Otherwise they are
zero in the nonrelativistic limit due to the orthogonality of the
radial wave functions. Then, assuming $\omega_{if}\sim m(\alpha
Z)^2$ r.u., $\Delta E_{ni}\sim m(\alpha Z)^2$ r.u., and taking
into account the relation $\alpha^2{\,} \mbox{r.u.}=1{\,}
\mbox{a.u.}$ for the energy units, we arrive at the result
Eq.~(\ref{n35}).

The picture looks quite different in the "velocity" gauge. In this
case the contributions of the intermediate states with $n=1s$
(first term in Eq.~(\ref{n37})) and $n=2p$ (second term in
Eq.~(\ref{n37})) cancel fully with the contribution of the
interference term. This cancellation was checked numerically
within the accuracy of our numerical procedure. The remaining
positive-energy matrix elements of the magnetic dipole operator
are nonzero only with the introduction of the spin-orbit
interaction, i.e. are of the order $\displaystyle\langle
n||\mu||f\rangle\sim\frac{\sqrt{\alpha}}{m}(\alpha Z)^2$ a.u. Then
the total contribution of the positive-energy remainder is of the
order

\begin{equation}
\label{n371} \displaystyle W^{E1M1
(+)}_{2p1s}(\rm{velocity})\approx(\alpha Z)^{12} a.u.
\end{equation}

This seems to be in agreement with the estimate, given by
Drake~\cite{q11} for the E1M1 transition in two-electron ions in
the high $Z$ region ($Z\geq 27$) where the influence of the
interelectron interaction becomes less significant. However, we
have to remind that Eq.~(\ref{n371}) gives only the minor
contribution to the total two-photon E1M1 $2p-1s$ decay rate for
small $Z$ values in the "velocity" gauge. The major contribution
arises in this case from the negative-energy intermediate states
and scales like

\begin{equation}
\label{n39a} \displaystyle W^{E1M1
(-)}_{2p1s}(\rm{velocity})=W^{E1M1
(total)}_{2p1s}(\rm{velocity})=a^{E1M1}_{2p1s}(\alpha Z)^8 a.u.
\end{equation}
with the same $a^{E1M1}_{2p1s}$ coefficients as in
Eq.(\ref{n35a}). The scaling behavior in Eq.(\ref{n39a}) will be
demonstrated explicitly in Section 7.

\begin{center}
\textbf{6. $E1E2$ transition probabilities.}
\end{center}
The E1E2 transition probabilities by the order of magnitude are
comparable with the E1M1 transition probabilities for all $Z$
values (see Table 3). This transition probability was also
evaluated in the "velocity" and "length" gauge within the same
numerical approach.

For the small Z values the $E^{E1E2}$ transition probability
scales with Z in the same way, as $W^{E1M1}$:
\begin{equation}
\label{n40} \displaystyle W^{E1E2}_{2p1s}=a^{E1E2}_{2p1s}(\alpha
Z)^8 a.u.
\end{equation}
\begin{equation}
\label{n41} \mbox{with} {\phantom{qqq}}\displaystyle
a^{E1E2}_{2p1s}=1.986\cdot10^{-5}
\end{equation}

In the "length" gauge exclusively the positive-energy intermediate
states contribute to the result Eq.~(\ref{n40}). The scaling law
for this contribution follows from the estimate similar to
Eq.~(\ref{n37}):

\begin{equation}
\label{n42} \displaystyle
W^{E1E2(+)}_{2p1s}\sim\frac{\alpha^2}{\pi}\omega^9_{if}\left|\sum_{n(+)}\left\{\frac{\langle
i||Q_{20}||n\rangle\langle n||d||f\rangle}{\Delta
E_{ni}}+\frac{\langle i||d||n\rangle\langle
n||Q_{20}||f\rangle}{\Delta E_{ni}} \right\}\right|^2
\end{equation}
where $Q_{20}$ is the spherical component of the quadrupole
electric transition operator $\displaystyle
Q_{2m}=\sqrt{\frac{4\pi\alpha}{5}}r^2Y_{2m}(\Omega)$. Here
$Y_{2m}$ is the spherical function, dependent on the electron
angular variables. The matrix elements of $Q$ in Eq.(\ref{n42})
are of the order $\langle
i||Q||n\rangle\sim\sqrt{\alpha}/{m(\alpha Z)^2}$. Inserting this
estimate in Eq.(\ref{n42}) we arrive at the result Eq.(\ref{n40}).

In the "velocity" gauge all the contribution from the
positive-energy, negative-energy and intermediate parts are
comparable. The value and the scaling behavior of the
negative-energy part will be derived analytically in Section 7.

\begin{center}
\textbf{7.Analytic derivation of the negative-energy contribution
to the E1M1 and E1E2 $2p_{1/2}-1s_{1/2}$ transition probabilities
in the "velocity" gauge for small $Z$ values.}
\end{center}
In this section we derive an explicit expression for the
negative-energy contribution to the E1M1 and E1E2
$2p{1/2}-1s_{1/2}$ transition probabilities in the "velocity"
gauge for small $Z$ values. This derivation will help us to check
the validity of our numerical calculations in the nonrelativistic
region (small $Z$ values).

We will perform this derivation using another set of the photon's
characteristics, namely photon's momentum \textbf{k} and
polarization vector \textbf{e}. Thus we will not be able to
distinguish between E1M1 and E1E2 transition probabilities and
will evaluate them as a unique correction to the dominant E1
$2p_{1/2}-1s_{1/2}$ transition.

The starting point for our calculations is the formula (\ref{u13})
where we retain the sum only over the negative-energy states. The
corresponding energy denominators in the nonrelativistic regime we
replace by -2m, neglecting also the photon frequencies, limited by
the value $\omega_{if}=E_A-E_{A'}=E_{2p_{1/2}}-E_{2s_{1/2}}$. Here
m is the electron mass. We also expand both exponents
Eq.(\ref{u28}) in Eq.(\ref{u13}), replacing one of these exponents
by the unity and retaining omly the next term of the expansion in
another exponent. After the summation over photon's polarizations
and the integration over the photon's emission directions this
will give us the desired correction to the leading E1
$2p_{1/2}-1s_{1/2}$ transition amplitude.

Thus we start with the expression
\begin{equation}
\label{n43} \displaystyle
W^{(-)}_{if}(\omega\omega')=\frac{\alpha^2}{(2\pi)^3}\omega\omega'\int
d{\bm{\nu}}{\bm{\nu}'}\sum_{{\textbf{e}}{\textbf{e}'}}|\tilde{U}_{if}^{(-)}|^2d\omega
\end{equation}
where
\begin{equation}
\label{n44} \displaystyle
\tilde{U}^{(-)}_{if}=-\frac{i}{2m}\left\{\langle
i|(\textbf{e}{\bm{\sigma}})(\textbf{k}\textbf{r})|n_{(-)}\rangle
\langle n_{(-)}|(\textbf{e}'{\bm{\sigma}})|f\rangle + \langle
i|(\textbf{e}{\bm{\sigma}})|n_{(-)}\rangle \langle
n_{(-)}|(\textbf{e}'{\bm{\sigma}})(\textbf{k}'\textbf{r})|n_{(-)}\rangle+[\textbf{e},\textbf{k}\leftrightarrow
\textbf{e}',\textbf{k}']\right\}
\end{equation}

Eq.(\ref{n44}) corresponds to the "velocity" gauge. Here
${\bm{\sigma}}$ are the Pauli matrices and the summation is
extended over the negative-energy set of solutions of the Dirac
equation for the electron in the field of the nucleus. In the
nonrelativistic limit this set comes over to the complete set of
the solutions of the Schr\"{o}dinger equation for the positron in
the field of the nucleus.

The employment of the closure relation then yields

\begin{equation}
\label{n45} \displaystyle
\tilde{U}^{(-)}_{if}=-\frac{i(\textbf{e}\textbf{e}')}{m}\langle
i|[(\textbf{k}\textbf{r})+(\textbf{k}'\textbf{r}')]|f\rangle
\end{equation}

For the summation over the polarizations we apply the standard
formula~\cite{q17}
\begin{equation}
\label{n46} \displaystyle
\sum_{\textbf{e}}e_ke_i=\delta_{ki}-\nu_k\nu_i
\end{equation}
Then
\begin{equation}
\label{n47} \displaystyle
\sum_{\textbf{e}\textbf{e}'}(\textbf{e}\textbf{e}')^*(\textbf{e}\textbf{e}')=2-({\bm{\nu}}{\bm{\nu}'})^2
\end{equation}
For the integration over the photon's emission angles the
following relations can be used:
\begin{equation}
\label{n48} \displaystyle \int d{\bm{\nu}}\nu_i=\int
d{\bm{\nu}}\nu_i\nu_k\nu_l=0
\end{equation}
\begin{equation}
\label{n49} \displaystyle \int
d{\bm{\nu}}\nu_i\nu_k=\frac{4\pi}{3}\delta_{ik}
\end{equation}
\begin{equation}
\label{n50} \displaystyle \int
d{\bm{\nu}}\nu_i\nu_j\nu_k\nu_l=\frac{4\pi}{15}(\delta_{ij}\delta_{kl}+\delta_{ik}\delta_{jl}+\delta_{il}\delta_{jk})
\end{equation}
The summation over the polarization and the integration over the
photon's emission angles in Eq.(\ref{n43}) yields
\begin{equation}
\label{n51} \displaystyle \int
d{\bm{\nu}}d{\bm{\nu}'}(k_i+k_i')(k_j+k_j')[2-({\bm{\nu}}\times{\bm{\nu}'})^2]=(\omega^2+\omega'^2)\frac{64\pi^2}{9}
\delta_{ij}
\end{equation}
Then, inserting Eq.(\ref{n51}) in Eq.(\ref{n43}) and using
Eq.(\ref{fw1}) we obtain
\begin{equation}
\label{n52} \displaystyle
W_{if}^{(-)}=\frac{4\alpha^2}{9m^2\pi}|\langle
i|\textbf{r}|f\rangle|^2\int_{0}^{\omega_{if}}\omega(\omega_{if}-\omega)[\omega^2+(\omega_{if}-\omega^2)^2]d\omega=
\frac{2\alpha^2}{45m^2\pi}\omega^5_{if}|\langle
i|\textbf{r}|f\rangle|^2
\end{equation}
The order of magnitude and the scaling behavior for the
transitions
\begin{equation}
\label{n53} \displaystyle
W_{2p1s}^{E1M1(-)}(\rm{velocity})+W_{2p1s}^{E1E2(-)}(\rm{velocity})=a^{(-)nr}_{2p1s}(\alpha
Z)^8 a.u.
\end{equation}
with
\begin{equation}
\label{n54} \displaystyle a^{(-)nr}_{2p1s}=5.625\cdot10^{-5}
\end{equation}
follow immediately from Eq.(\ref{n52})in the nonrelativistic
limit. This result can be compared with the exact result of the
numerical evaluations:
\begin{equation}
\label{n55} \displaystyle
W_{2p1s}^{E1M1(-)}(\rm{velocity})+W_{2p1s}^{E1E2(-)}(\rm{velocity})=a^{(-)}_{2p1s}(\alpha
Z)^8 a.u.
\end{equation}
with
\begin{equation}
\label{n56} \displaystyle a^{(-)}_{2p1s}=5.806\cdot10^{-5}
\end{equation}
The discrepancy between $a^{(-)}_{2p1s}$ and $a^{(-)nr}_{2p1s}$
(3,1\%) is larger than the expected discrepancy due to the
relativistic corrections
\begin{equation}
\label{n57}
\frac{(a^{(-)}_{2p1s}-a^{(-)nr}_{2p1s})}{a^{(-)}_{2p1s}}\simeq(\alpha
Z)^2
\end{equation}
for Z=1. Besides, the discrepancy in our case does not depend on
$Z$ for small $Z$ values. It can be argued, however, that the
order of magnitude and Z-dependence (\ref{n57}) correspond to the
positive-energy contributions and can be different in case of the
negative-energy intermediate states.

In total, the discrepancy between $a^{(-)nr}_{2p1s}$ and
$a^{(-)}_{2p1s}$ is small enough to confirm the validity of our
numerical calculations in the region of the small $Z$ values.

\begin{center}
\textbf{8. Conclusions.}
\end{center}
In this work we have evaluated for the first time the two-photon
transitions probabilities $W^{E1M1}_{2p1s}$ and $W^{E1E2}_{2p1s}$
in the H atom and in H-like ions with $Z$ up to $Z=100$. The
evaluation is performed in a fully relativistic way. For the small
$Z$ values the scaling law $(\alpha Z)^8$ is established for both
probabilities. The total transition probability for $2p\rightarrow
1s$ transition can be presented in a form
\begin{equation}
\label{n58} \displaystyle W=W_0\left[1+\frac{\alpha}{\pi}(\alpha
Z)^4F(\alpha Z) \right] a.u.
\end{equation}
where $W_0$ is the transition probability for $2p\rightarrow
\gamma(E1)+1s$ process:
\begin{equation}
\label{n59} \displaystyle
W_0=\frac{4}{3}\omega_{if}^3\left|\langle
i|\textbf{r}|f\rangle\right|^2\alpha^3 a.u.
\end{equation}

The function $F(\alpha Z)$ for all $Z$ is given in Table 4. For
$Z=1$ the correction term in Eq.(\ref{n58}) is quite small, but
grows very fast with the increase of the nuclear charge $Z$ and
becomes significant in the one-electron highly charged ions.

\begin{center}
\textbf{Acknowledgements.}
\end{center}
This work was supported by the RFBR grant $N^{o} 02-02-16578$ and
by Minobrazovanje grant E02.-3.1-7.

\newpage
Table 1. Total two-photon $E1E1$ $2s_{1/2}-1s_{1/2}$ transition
probabilities in $sec^{-1}$  for different Z values. Numbers in
parentheses are the powers of 10.

\begin{tabular}{||c c c c||c c c c||}
\hline \hline
Z & $W^{(\text{vel})}$ & $W^{(\text{len})}$ &$W^{\text{~\cite{q5}}}$ & Z &$W^{(\text{vel})}$ &$W^{(\text{len})}$ & $W^{\text{~\cite{q5}}}$\\
\hline {}
1 & 8.2205 & 8.2207& 8.2291 & 40 & 3.1954(10) & 3.1956(10) & 3.1990(10)  \\
2 & 5.2605(2)& 5.2607(2) & 5.2661(2) & 45 & 6.3927(10) & 6.3926(10) & 6.4003(10)  \\
3 & 5.9909(3)& 5.9910(3) & 5.9973(3) & 50 & 1.1854(11) & 1.1854(11) & 1.1869(11)  \\
4 & 3.3652(4)& 3.3653(4) &3.3689(4) & 56 & 2.2947(11)& 2.2948(11)&  2.2980(11) \\
5 & 1.2833(5)& 1.2834(5) &1.2847(5) & 60 & 3.4229(11) & 3.4230(11) & 3.4282(11)  \\
6 & 3.8305(5)& 3.8306(5) &3.8347(5) & 65 & 5.4293(11) & 5.4293(11) & 5.4387(11)  \\
7 & 9.6550(5)& 9.6551(5) &9.6654(5) & 70 & 8.2975(11) & 8.2975(11) & 8.3139(11)  \\
8 & 2.1500(6)& 2.1502(6) &2.1525(6) & 74 & 1.1379(12) & 1.1379(12) & 1.1404(12)  \\
9 & 4.3564(6)& 4.3565(6) &4.3612(6) & 80 & 1.7655(12) &1.7655(12) & 1.7701(12)  \\
10 & 8.1921(6)& 8.1922(6)&8.2010(6) & 85 & 2.4747(12) & 2.4748(12) & 2.4824(12)  \\
12 & 2.4425(7)& 2.4425(7) &2.4451(7) & 90 & 3.3899(12) &3.3899(12) & 3.4021(12)  \\
15 & 9.2914(7)& 9.2915(7) &9.3017(7) & 92 & 3.8216(12) &3.8216(12) & 3.8361(12)  \\
20 & 5.1898(8)& 5.1899(8) &5.1956(8) & 100 & 5.9782(12) & 5.9783(12)& 6.0045(12)  \\
30 & 5.8151(9)& 5.8152(9) &5.8217(9) &  &  & &\\
\hline \hline
\end{tabular}

\newpage
Table 2. The two-photon E1M1 $2p_{1/2}-1s_{1/2}$ transition
probabilities in $sec^{-1}$ for different $Z$ values in the
"velocity" and "length" gauge. The positive-energy $W_+$,
negative-energy $W_-$ and total contributions $W_t$ are given.

\begin{tabular}{||c| c c c| c c c ||}
\hline \hline
Z & & velocity & & &length &\\
\hline {}
 &$W_+$ & $W_-$ & $W_t$& $W_+$& $W_-$& $W_t$ \\
1 & 3.223(-5) & 9.667(-6) & 9.667(-6) & 9.667(-6) &5.889(-16) & 9.667(-6)\\
2 & 8.249(-3)  & 2.475(-3)  & 2.474(-3) & 2.474(-3)&2.410(-12) &  2.474(-3)\\
4 & 2.111  & 6.334(-1)  & 6.332(-1) & 6.331(-1) &9.886(-9) &  6.331(-1)\\
8 & 5.407(2)  & 1.625(2)  & 1.619(2) & 1.619(2) &4.056(-5) & 1.619(2) \\
10 & 3.219(3) & 9.702(2) & 9.637(2) & 9.637(2) &5.910(-4) & 9.637(2) \\
14 & 4.745(4) & 1.437(4)  & 1.418(4) & 1.418(4) & 3.361(-2)& 1.418(4) \\
18 & 3.538(5) & 1.078(5) & 1.054(5) & 1.051(5) & 6.888(-1)& 1.055(5) \\
20 & 8.211(5) & 2.510(5) & 2.443(5) & 2.433(5) &2.445 & 2.443(5) \\
24 & 3.523(6) & 1.086(6) & 1.045(6) & 1.039(6) & 2.193(1) & 1.045(6) \\
28 & 1.206(7) & 3.754(6)  & 3.562(6) &  3.533(5) &1.404(2) &  3.563(6)\\
30 & 2.092(7) & 6.547(6) & 6.164(6) & 6.105(6) &3.227(2) & 6.164(6) \\
34 & 5.674(7) & 1.798(7)  & 1.664(7) & 1.643(7) &1.462(3) & 1.664(7) \\
38 & 1.376(8) & 4.421(7)  & 4.014(7) & 3.951(7) &5.609(3) &  4.014(7)\\
40 & 2.070(8) &  6.700(7) & 6.020(7) & 5.914(7) & 1.044(4)&  6.021(7)\\
44 & 4.417(8) &  1.453(8) & 1.276(8) & 1.249(8) & 3.315(4)&  1.277(8)\\
48 & 8.812(8) & 2.952(8)  & 2.530(8) &  2.465(8)& 9.543(4)&  2.531(8)\\
50 & 1.218(9) &  4.119(8) & 3.486(8) & 3.386(8) & 1.569(5)&  3.486(8)\\
54 & 2.240(9) &  7.734(8) & 6.366(8) & 6.151(8) & 4.011(5)&  6.366(8)\\
58 & 3.939(9) & 1.391(9)  & 1.111(9) &  1.068(9)& 9.614(5)&  1.112(9)\\
64 & 8.549(9) & 3.134(9)  & 2.386(9) &  2.269(9)& 3.221(6)&  2.387(9)\\
68 & 1.375(10) & 5.180(9)  & 3.813(9) &  3.597(9)& 6.807(6)&  3.813(9)\\
70 & 1.725(10) &  6.593(9) & 4.767(9) &  4.479(9)& 9.744(6)&  4.767(9)\\
74 & 2.659(10) & 1.048(10)  & 7.307(9) &  6.805(9)& 1.943(7)&  7.308(9)\\
78 & 3.999(10) & 1.629(10)  & 1.094(10) &  1.009(10)& 3.747(7)&  1.094(10)\\
80 & 4.863(10) & 2.016(10)  & 1.327(10) &  1.218(10)& 5.145(7)&  1.328(10)\\
84 & 7.075(10) &  3.042(10) & 1.927(10) &  1.749(10)& 9.502(7)&  1.927(10)\\
88 & 1.009(11) & 4.511(10)  & 2.748(10) &  2.466(10)& 1.711(8)&  2.748(10)\\
90 & 1.196(11) & 5.460(10)  & 3.262(10) & 2.910(10) & 2.276(8)&  3.262(10)\\
92 & 1.411(11) & 6.583(10) & 3.859(10) & 3.422(10) & 3.012(8)& 3.859(10) \\
94 & 1.658(11) & 7.910(10)  & 4.550(10) & 4.010(10) & 3.967(8)&  4.551(10)\\
98 & 2.260(11) & 1.130(11)  & 6.269(10) & 5.456(10) & 6.781(8)&  6.270(10)\\
100 & 2.621(11) & 1.344(11) & 7.329(10) & 6.339(10) &8.805(8) & 7.330(10)\\
\hline \hline
\end{tabular}

\newpage

\newpage

Table 3. The two-photon E1E2 $2p_{1/2}-1s_{1/2}$ transition
probabilities in $sec^{-1}$ for different $Z$ values in the
"velocity" and "length" gauge. The positive-energy $W_+$,
negative-energy $W_-$ and total contributions $W_t$ are given.

\begin{tabular}{||c| c c c| c c c ||}
\hline
Z & & velocity & & &length &\\
\hline {}
 &$W_+$ & $W_-$ & $W_t$& $W_+$& $W_-$& $W_t$ \\
1 & 1.232(-6) & 9.667(-6)& 6.605(-6) & 6.604(-6) &3.625(-27) & 6.604(-6)\\
2 & 3.154(-3) & 2.474(-3)& 1.690(-3) & 1.691(-3) & 2.375(-22)& 1.691(-3) \\
4 & 8.075(-2) &6.335(-1) & 4.327(-1) & 4.326(-1) & 1.557(-17)& 4.326(-1) \\
8 & 2.065(1) &1.620(2) & 1.105(2) & 1.106(2) &1.023(-12) & 1.106(2) \\
10 & 1.230(2) &  9.655(2)& 6.583(2) & 6.582(2) &3.638(-11) &6.582(2)  \\
14 & 1.813(3) &  1.423(4)& 9.683(3) & 9.681(3) & 7.949(-9)& 9.681(3) \\
18 & 1.351(4) &  1.061(5)&7.197(4)  & 7.196(4) & 4.452(-7)& 7.196(4) \\
20 & 3.136(4) &2.463(5)  &1.668(5)  & 1.667(5) & 2.409(-6)& 1.667(5) \\
24 & 1.345(5) & 1.057(6) & 7.125(5) & 7.125(5) &4.482(-5) & 7.125(5) \\
28 & 4.605(5) & 3.620(6) & 2.428(6) & 2.427(6) & 5.321(-4)& 2.427(6) \\
30 & 7.987(5) & 6.278(6) & 4.198(6) & 4.198(6) & 1.611(-3)& 4.198(6) \\
34 & 2.167(6) & 1.703(7)  & 1.132(7) & 1.132(7) & 1.206(-2)& 1.132(7) \\
38 & 5.262(6) & 1.135(7)  & 2.728(7) & 2.728(7) & 7.227(-2)& 2.727(7) \\
40 & 7.920(6) & 6.222(7) &4.088(7)  & 4.088(7) & 1.652(-1)& 4.088(7) \\
44 & 1.692(7) & 1.328(8)  & 8.654(7) & 8.655(7) & 7.696(-1)&  8.654(7)\\
48 & 3.386(7) & 2.654(8)  & 1.712(8) & 1.712(8) &3.144 &  1.712(8)\\
50 & 4.688(7) & 3.671(8)  & 2,356(8) & 2.356(8) &6.093 &  2.355(8)\\
54 & 8.658(7) & 6.762(8)  & 4.290(8) & 4.291(8) &2.126(1) &  4.290(8)\\
58 & 1.530(8) & 1.191(9)  & 7.465(8) & 7.468(8) &6.807(1) &  7.465(8)\\
64 & 3.361(8) & 2.596(9)  & 1.592(9) & 1.594(9) &3.404(2) &  1.592(9)\\
68 & 5.460(8) & 4.189(9)  & 2.530(9) & 2.532(9) & 9.217(2)&  2.530(9)\\
70 & 6.888(8) & 5.265(9)& 3.153(9) & 3.155(9) &1.486(3) &  3.152(9)\\
74 & 1.077(9) & 8.152(9)  & 4.793(9) & 4.801(9) & 3.726(3)& 4.793(9) \\
78 & 1.646(9) & 1.232(10)  & 7.103(9) & 7.115(9) & 8.941(3)& 7.103(9) \\
80 & 2.020(9) &1.502(10) & 8.568(9) & 8.589(9) & 1.364(4)& 8.569(9) \\
84 & 3.002(9) & 2.199(10) & 1.226(10) & 1.228(10) &3.092(4) &  1.226(10)\\
88 & 4.390(9) & 3.158(10) & 1.717(10) & 1.723(10) & 6.781(4)&  1.717(10)\\
90 & 5.280(9) & 3.759(10) &2.017(10)  & 2.025(10) &9.927(4) &  2.016(10)\\
92 & 6.329(9) &4.455(10) & 2.357(10) & 2.367(10) &1.444(5) &  2.356(10)\\
94 & 7.562(9) & 5.259(10)  & 2.740(10) & 2.754(10) & 2.085(5)&  2.740(10)\\
98 & 1.070(10) & 7.239(10)  & 3.653(10) & 3.677(10) & 4.272(5)&  3.653(10)\\
100 & 1.268(10) & 8.444(10) &4.187(10)  & 4.218(10) & 6.060(5)& 4.187(10)\\
\hline \hline
\end{tabular}

\newpage

Table 4. The function $F(\alpha Z)\cdot 10^{-3}$ for different $Z$
values.

\begin{tabular}{||c c||c c||}
\hline \hline
Z & $F(\alpha Z)$ & Z & $F(\alpha Z)$ \\
\hline {}
1  & 3.945& 50 & 3.569\\
2  &3.943 & 54 & 3.508\\
4  &3.942 & 58 & 3.443\\
8  &3.935 & 64 & 3.340\\
10 &3.929 & 68 & 3.268\\
14 &3.915 & 70 & 3.231\\
18 &3.894 & 74 & 3.153\\
20 &3.883 & 78 & 3.075\\
24 &3.857 & 80 & 3.034\\
28 &3.825 & 84 & 2.954\\
30 &3.807 & 88 & 2.873\\
34 &3.768 & 90 & 2.833\\
38 &3.725 & 92 & 2.794\\
40 &3.702 & 94 & 2.754\\
44 &3.651 & 98 & 2.677\\
48 &3.597 & 100 & 2.640\\
\hline \hline
\end{tabular}

\newpage

\begin{figure}[h]
\hfil\epsffile{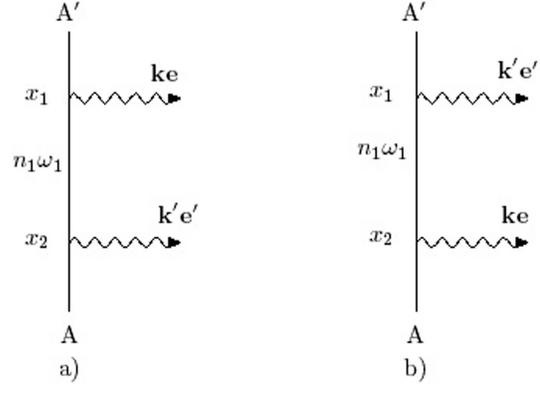}\hfil \caption{The Feynman
graphs,corresponding to the two-photon decay process $A\to A'$.
The photons are characterized by the momentum $\textbf{k}$ and the
polarization $\textbf{e}$}\label{qq1}
\end{figure}
\newpage

\end{document}